\documentclass[pra,aps,showpacs,superscriptaddress,twocolumn]{revtex4}

\usepackage{amsmath}
\usepackage{amsfonts}
\usepackage{bm}
\usepackage{graphicx}
\usepackage{array,color}

\begin{document}
\title{Twisted spin vortices in a spinor-dipolar Bose-Einstein condensate with Rashba spin-orbit coupling}

\author{Masaya Kato}
\affiliation{Department of Engineering Science, University of Electro-Communications, Tokyo 182-8585, Japan}

\author{Xiao-Fei Zhang}
\affiliation{Department of Engineering Science, University of Electro-Communications, Tokyo 182-8585, Japan}
\affiliation {Key Laboratory of Time and Frequency Primary Standards, National Time Service Center, Chinese Academy of Sciences, Xi'an 710600, China}

\author{Daichi Sasaki}
\affiliation{Department of Engineering Science, University of Electro-Communications, Tokyo 182-8585, Japan}

\author{Hiroki Saito}
\affiliation{Department of Engineering Science, University of Electro-Communications, Tokyo 182-8585, Japan}

\date{\today}

\begin{abstract}

We consider a spin-1 Bose-Einstein condensate with Rashba spin-orbit
coupling and dipole-dipole interaction confined in a cigar-shaped trap.
Due to the combined effects of spin-orbit coupling, dipole-dipole
interaction, and trap geometry, the system exhibits a rich variety of
ground-state spin structures, including twisted spin vortices.
The ground-state phase diagram is determined with respect to the strengths
of the spin-orbit coupling and dipole-dipole interaction.
\end{abstract}

\pacs{03.75.Mn, 03.75.Lm, 67.85.Bc, 67.85.Fg}

\maketitle

\section{Introduction}

Spin-orbit coupling is of fundamental importance in many branches of
physics, such as quantum spin-Hall effect, topological insulators, and
superconductivity~\cite{D. C. Tsui,I. Zutic,M. Z. Hasan,X. L. Qi}.
Recently, the NIST group has realized the light-induced vector potentials
and the synthetic electric and magnetic fields in Bose-Einstein
condensates (BECs) of neutral atoms using Raman processes
\cite{Lin01,Y.-J. Lin1,Lin11,J. Dalibard}.
Remarkably, they also created a two-component spin-orbit coupled
condensate of Rb atoms \cite{Lin02}.
Artificial spin-orbit coupling (SOC) offers us a tremendous opportunity to
study exotic quantum phenomena in many-body systems, which exhibit
various symmetry-broken and topological condensate
phases in pseudospin-$1/2$
systems~\cite{C. J. Wu,H. Zhai0,C. Wang,T.-L. Ho,X.-Q. Xu,S. Sinha,H. Hu,B. Ramachandhran,T. Kawakami}. 
For spin-$1$ and -$2$ condensates, more exotic patterns form due to the
competition between the SOC and spin-dependent interactions
\cite{Z. F. Xu,T. Kawakami1,S.-W. Su,C.-F. Liu,Z. Chen}.

On the other hand, recent experimental realization of BECs of atomic
species with large magnetic moments boosts interest in the field of
quantum gases with dipole-dipole interaction
(DDI)~\cite{Gries,T. Lahaye,M. Lu,K. Aikawa}.
Previous studies on spinor-dipolar BECs have shown that the interplay
between spin-dependent interaction and DDI leads to rich topological
defects and spin structures
\cite{Kawaguchi1,Yi,Kawaguchi2,Kawaguchi3,Gaw07,Helical,Eto}.
Consequently, it is of particular interest to explore the effects of
long-rang and anisotropic DDI on such a spin-orbit coupled system,
which has recently drawn considerable attentions.
More specifically, Deng {\it et al}.~\cite{Y. Deng} proposed an
experimental scheme to create SOC in spin-3 Cr atoms using Raman
processes.
Wilson {\it et al}.\ have investigated the effects of DDI on a
pseudospin-$1/2$ spin-orbit coupled condensate, and predicted the
emergence of a thermodynamically stable ground state having a spin
configuration called meron~\cite{Meron}.
Furthermore, a number of quantum crystalline and quasicrystalline ground
states were found in two-dimensional (2D) dipolar bosons with Rashba
SOC~\cite{Quantum Quasicrystals}.

In this work, we consider a BEC of spin-$1$ bosons confined in a
cigar-shaped trap potential, subject to both 2D SOC and DDI.
The 2D SOC tends to create spin textures in the $x$-$y$ plane, while the
DDI can generate $z$-dependent spin textures in an elongated system.
As a result, 3D spin structures emerge in this system.
We elucidate the ground-state spin textures as functions of the strengths
of the SOC and DDI by numerically minimizing the energy functional.
We will show a rich variety of ground-state spin textures, such as twisted
spin vortices, in which spin vortices twist around each other along the
$z$ direction.

The paper is organized as follows.
In Sec.~\ref{s:formulation}, we formulate the theoretical model and
briefly introduce the numerical method.
In Sec.~\ref{s:result}, the ground-state phase diagram of the system is
determined, and a detailed description of each phase is given.
In Sec.~IV, the main results of the paper are summarized.

\section{Formulation of the problem}
\label{s:formulation}

We consider a BEC of spin-$1$ atoms with mass $M$ confined in a harmonic
potential, which are subject to the 2D SOC.
We employ the mean-field approximation and the state of the system is
described by the spinor order parameter $\bm\Psi(\bm r)=(\psi_{1}(\bm r),
\psi_{0}(\bm r),\psi_{-1}(\bm r))^{\rm T}.$
The single-particle energy is given by
\begin{equation}
\label{eq:H0}
E_{0}= \int d \bm r \bm\Psi^\dagger \left[ -\frac{\hbar^2}{2M} \nabla^2 +
V(\bm r) + \textmd{g}_{\rm soc} \frac{\hbar}{i} \nabla_\perp \cdot
\bm{f}_\perp \right] \bm\Psi,
\end{equation}
where $\textmd{g}_{\rm soc}$ parametrizes the SOC strength, $\nabla_\perp
= (\partial_x, \partial_y)$, and $\bm{f}_\perp = (f_x, f_y)$ are the $3
\times 3$ spin-$1$ matrices.
The trap potential is axisymmetric, $V(\bm r) = M \omega_\perp^2
(x^2+y^2+\lambda^2z^2)/2$, where $\omega_\perp$ is the radial trap
frequency and $\lambda = \omega_z / \omega_\perp$ is the aspect ratio
between the axial and radial trap frequencies.
The $s$-wave contact interaction energy is written as
\begin{equation} \label{eq:Hint}
E_s = \frac{1}{2}\int d \bm r\left[\textmd{g}_0 \rho(\bm r) +
\textmd{g}_1 \bm F^2(\bm r) \right],
\end{equation}
where $\textmd{g}_0 = 4\pi \hbar^2 (a_0+2a_2)/ (3 M)$ and $\textmd{g}_1 =
4\pi \hbar^2 (a_0-a_2)/ (3 M)$ with $a_{\rm s} (s=0, 2)$ being the
$s$-wave scattering length for the scattering channel with total spin $s$.
The total atomic density $\rho(\bm r)=|\psi_{1}(\bm r)|^2+|\psi_{0}(\bm
r)|^2 + |\psi_{-1}(\bm r)|^2$ satisfies $\int \rho(\bm r) d\bm{r} =
N$, where $N$ is the total number of atoms.
The spin density has the form
\begin{equation}
\label{eq:spin}
\bm F(\bm r) = \bm\Psi^\dagger
\begin{bmatrix}
f_x\\
f_y\\
f_z
\end{bmatrix}
\bm\Psi
=
\begin{bmatrix}
\sqrt 2\rm{Re}[\psi^\ast_1\psi_0 + \psi^\ast_0\psi_{-1}]\\
\sqrt 2\rm{Im}[\psi^\ast_1\psi_0 + \psi^\ast_0\psi_{-1}]\\
|\psi_1|^2-|\psi_{-1}|^2
\end{bmatrix}.
\end{equation}
The DDI energy is given by
\begin{equation}
E_{\rm ddi}=\frac{\textmd{g}_{\rm dd}}{2}\int d\bm rd\bm r'
\frac{\hat{\bm F}(\bm r)\cdot \hat{\bm F}(\bm r')
-3(\hat{\bm F}(\bm r)\cdot\bm e)(\hat{\bm F}(\bm r')\cdot\bm e)}{|\bm r-\bm r'|^3}, \label{eq:Eddi}
\end{equation}
where $\textmd{g}_{\rm dd}=\mu_0\mu^{2}_{\rm d} / (4\pi)$, $\mu_0$ is the
magnetic permeability of the vacuum, $\mu_{\rm d}$ is the magnetic dipole
moment of the atom, and $\bm e=(\bm r -\bm r')/|\bm r - \bm r'|$.
The total energy of the system is thus given by
$E = E_0 + E_s + E_{\rm ddi}$. 

The ground state is numerically obtained by minimizing the total energy
$E$ using the imaginary-time propagation method.
For the imaginary-time evolution, the pseudospectral method with the
fourth-order Runge-Kutta scheme is used.
In the following numerical simulations, we work in dimensionless unit.
The energy and length are normalized by $\hbar\omega_{\perp}$ and
$a_{\perp} = \sqrt{\hbar/(M\omega_{\perp})}$.
In this unit, the wave function, the SOC coefficient
$\textmd{g}_{\rm soc}$, and the interaction coefficients $\textmd{g}_0$,
$\textmd{g}_1$, and $\textmd{g}_{\rm dd}$ are normalized by
$N^{1/2} / a_\perp^{3/2}$, $a_\perp \omega_\perp$, and $\hbar \omega_\perp
a_\perp^3 / N$, respectively.

\section{Ground-state spin structures}
\label{s:result}

The richness of the present system lies in the large number of free
parameters, including the strength and sign of the contact interactions,
DDI, SOC, aspect ratio, and so on.
To highlight the effects of the SOC and DDI, we fix $\lambda=0.2$,
$\textmd{g}_0=4000$, and $\textmd{g}_1=0$, implicitly assuming that the
ground-state spin texture is dominated by the SOC and DDI.

\begin{figure}[tb]
\includegraphics[width=8.0cm]{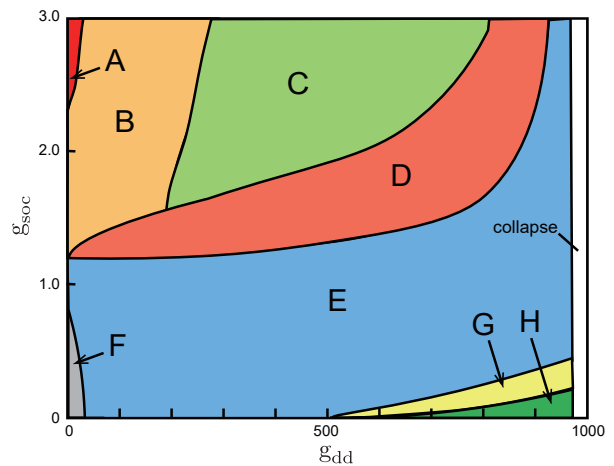}
\caption{(color online) Ground-state phase diagram of the spin-orbit
coupled dipolar BEC with respect to $\textmd{g}_{\rm soc}$ and
$\textmd{g}_{\rm dd}$ for $\textmd{g}_0=4000$, $\textmd{g}_1=0$, and
$\lambda=0.2$.
There are eight different phases marked by A-H.
The white region represents instability against dipolar collapse.
}
\label{f:PD}
\end{figure}
Our main results are summarized in Fig.~\ref{f:PD}, which shows the
ground-state phase diagram of a spin-orbit coupled dipolar condensate with
respect to $\textmd{g}_{\rm soc}$ and $\textmd{g}_{\rm dd}$.
There are eight different phases marked by A-H, which differ in density
profiles, spin texture and angular momentum.
In the following discussion, we will give a detailed description of each
phase.
In the white region of Fig.~\ref{f:PD}, the condensate
collapses due to the attractive part of the DDI~\cite{Lahaye}, where no
stable mean-field solution exists~\cite{Koch}.
The critical value of $\textmd{g}_{\rm dd}$ for the collapse seems almost
independent of $\textmd{g}_{\rm soc}$.

We start from the case where both the SOC and DDI are sufficiently weak,
indicated by the gray region F in Fig. \ref{f:PD}.
In this phase, the central region of the potential is occupied by $m_f=0$
component and the system is condensed to such component, leading to
vanishing magnetization of the system.
We note that this phase disappears with increasing either the SOC or DDI.

\begin{figure}[tb]
\includegraphics[width=9cm]{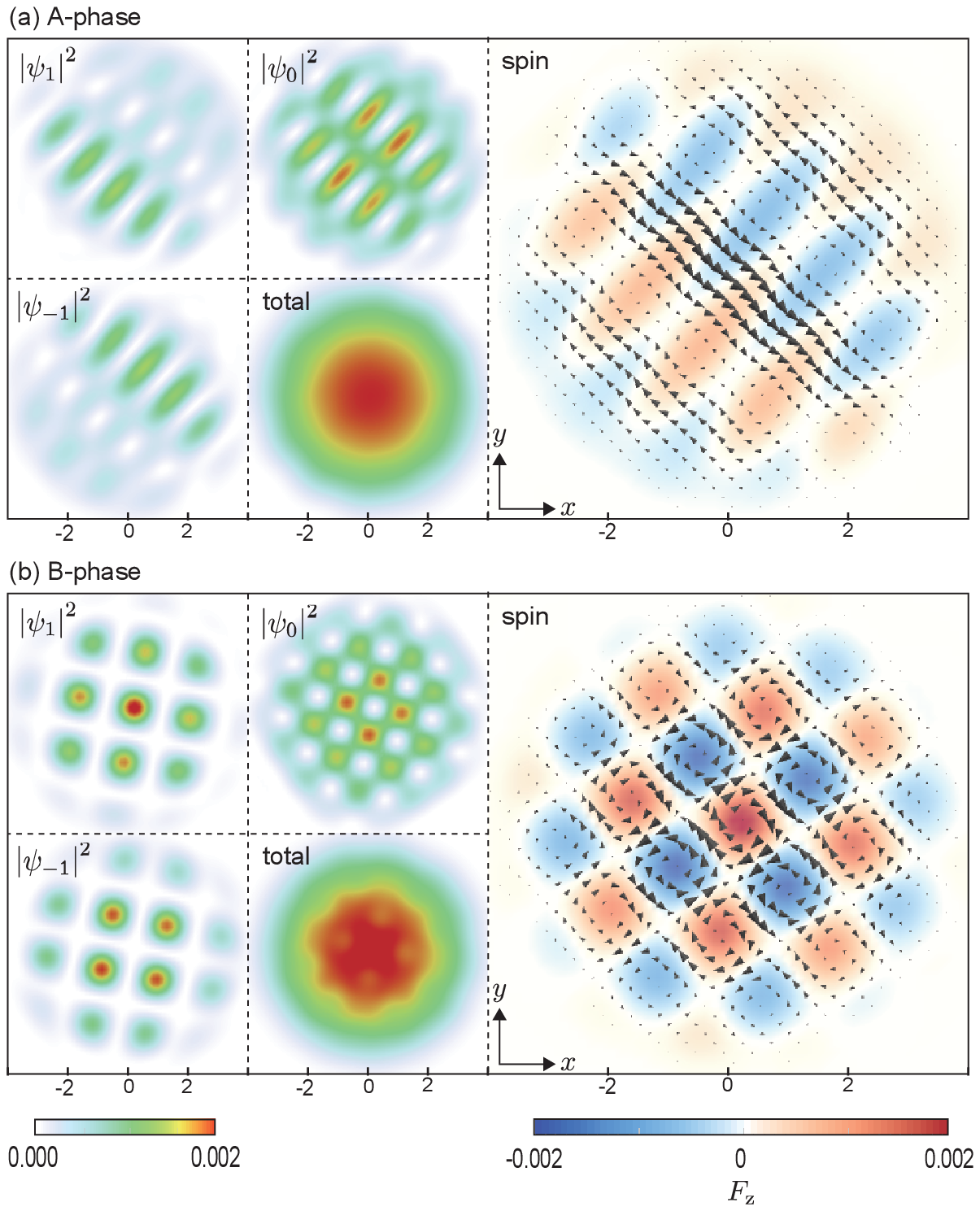}
\caption{(color online) Typical density distribution and spin texture of
the system for
(a) $\textmd{g}_{\rm dd}=0.0$ and $\textmd{g}_{\rm soc}=2.4$ and
(b) $\textmd{g}_{\rm dd}=100$ and $\textmd{g}_{\rm soc}=2.4$,
corresponding to the states represented in
A and B phases in Fig. \ref{f:PD}, respectively.
The arrows in the spin texture represent the transverse spin vector
$(F_{\rm x}, F_{\rm y})$ with background color representing $F_{\rm z}$.}
\label{f:SOC}
\end{figure}
In the limit of strong SOC but weak DDI, the system exhibits a spin-stripe
pattern, indicated by A-phase in Fig.~\ref{f:PD}.
Typical density and spin distributions of such phase are shown in
Fig.~\ref{f:SOC}(a).
In this phase, the spin texture on the $x$-$y$ plane shows typical spin
stripe structure, which is almost unchanged along the $z$-axis.
Actually, previous studies on trapped spin-orbit coupled BECs have shown
that the spin stripe structure is known as one of the ground states at
strong SOC in harmonic potential~\cite{S. Sinha}.
In the present system, this state also exists for a strong SOC, but with a
weak DDI.

With an increase in the strength of the DDI, B-phase emerges as the ground
state, as shown in Fig. \ref{f:PD}. Typical density and spin distributions of such phase are shown in
Fig.~\ref{f:SOC}(b). This phase is characterized by the checkerboard lattice of spin vortices
on the $x$-$y$ plane, in which spin vortices with $F_z > 0$ and $F_z < 0$
are alternately aligned. Such a pattern may be understood by the long-range nature of
the DDI, which leads to a regular density distribution of each component.
Similar to A-phase, the spin texture is almost independent of $z$.

\begin{figure}[tb]
\includegraphics[width=9cm]{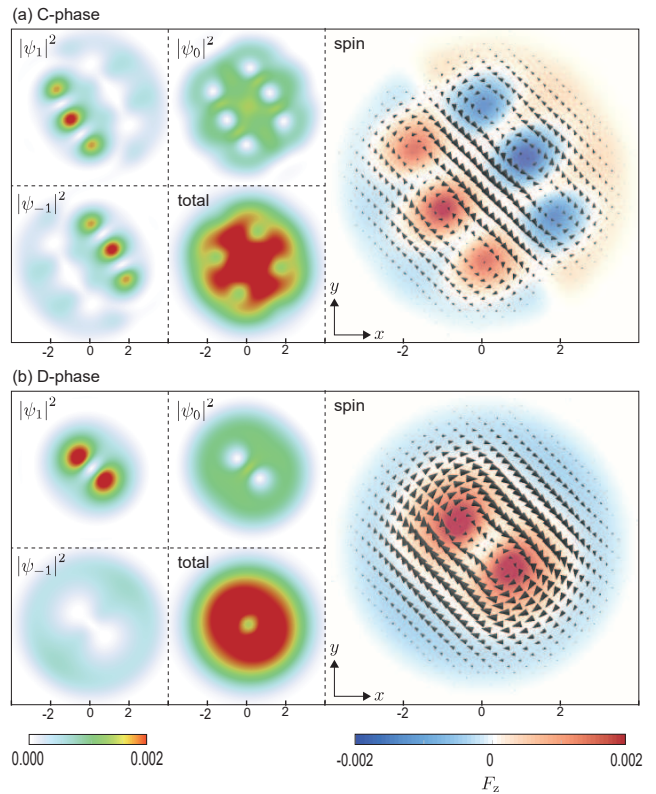}
\caption{(color online) Typical density distribution and spin texture of
the system for
(a) $\textmd{g}_{\rm dd}=250$ and $\textmd{g}_{\rm soc}=2.0$ and
(b) $\textmd{g}_{\rm dd}=300$ and $\textmd{g}_{\rm soc}=1.5$,
corresponding to C and D phases in Fig.~\ref{f:PD}, respectively.
The arrows in the spin texture represent the transverse spin vector
$(F_{\rm x}, F_{\rm y})$ with background color representing $F_{\rm z}$.
}
\label{f:SOCandDDI}
\end{figure}
\begin{figure}[tb]
\includegraphics[height=5.0cm,width=5.5cm]{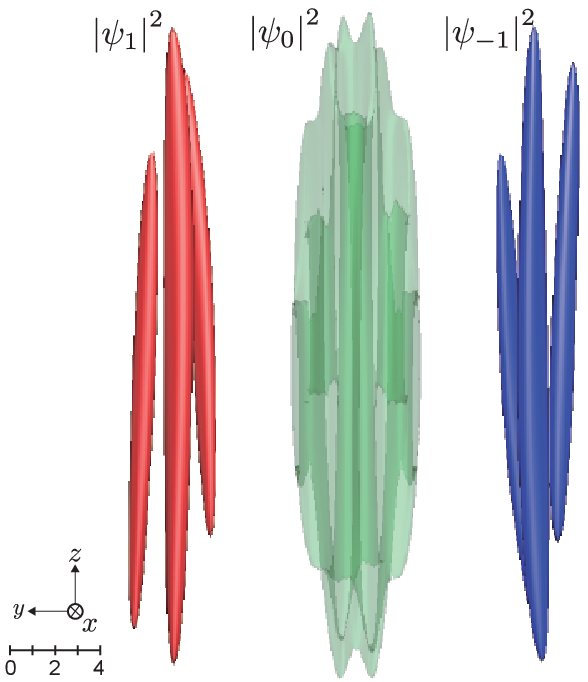}
\caption{(color online)
Isodensity surfaces of the three components
$|\psi_1|^2=0.001$ (red), $|\psi_{0}|^2=0.0007$ (green), and
$|\psi_{-1}|^2=0.001$ (blue), corresponding to C-phase shown in
Fig.~\ref{f:SOCandDDI}(a).
See the Supplemental Material for a movie showing the three-dimensional
(3D) structure~\cite{SM}.
} \label{f:C}
\end{figure}
Increasing the DDI further, the spin vortex structures begin to have a $z$
dependence, and C-phase emerges as the ground-state of the system, as the
yellow-green region in Fig. \ref{f:PD}.
Typical density and spin distributions of such phase are shown
in Fig.~\ref{f:SOCandDDI}(a).
This phase has a spin-vortex train structure on the $x$-$y$ plane, where
components $1$ and $-1$ are surrounded by component $0$.
The numbers of spin vortices with $F_{\rm z}>0$ and $F_{\rm z}<0$ are
equal to each other, which increase with SOC.
Three-dimensional (3D) isodensity surfaces of the state are shown in
Fig.~\ref{f:C}, which indicates that the spin vortex structure depends on
$z$ due to the DDI.

\begin{figure}[tb]
\includegraphics[height=5.0cm,width=5.5cm]{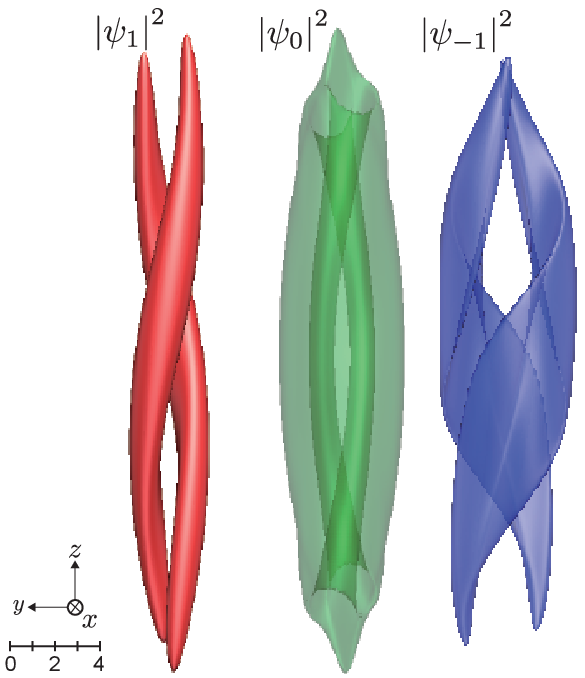}
\caption{(color online)
Isodensity surfaces of the three components
$|\psi_1|^2=0.001$ (red), $|\psi_0|^2=0.0007$ (green), and
$|\psi_{-1}|^2=0.0004$ (blue), corresponding to D-phase shown in
Fig.~\ref{f:SOCandDDI}(b).
See the Supplemental Material for a movie showing the 3D
structure~\cite{SM}.
} \label{f:D}
\end{figure}
With a further increase in the DDI, C-phase transforms to D-phase, as
shown in Fig.~\ref{f:PD}.
Its density and spin distributions and 3D structures are shown in
Figs.~\ref{f:SOCandDDI}(b) and \ref{f:D}, respectively.
Interestingly, the spin structure significantly depends on $z$ and form a
helical structure, leading to twisted spin vortices.
We note that this spin structure, as well as C-phase, reflects the
features of the SOC and DDI: multiple spin vortices are created by the SOC
and they are twisted along the $z$-axis by the DDI.
In the region D of Fig.~\ref{f:PD}, we also observe three twisted spin
vortices for a larger SOC.
Our numerical results show that the degree of torsion increases with DDI,
while the separation between the spin-vortices decreases.
In the limit of strong DDI, the separation almost disappears and E-phase
emerges as the ground state, as shown in blue region in Fig.~\ref{f:PD}.

\begin{figure}[tb]
 \includegraphics[width=9cm]{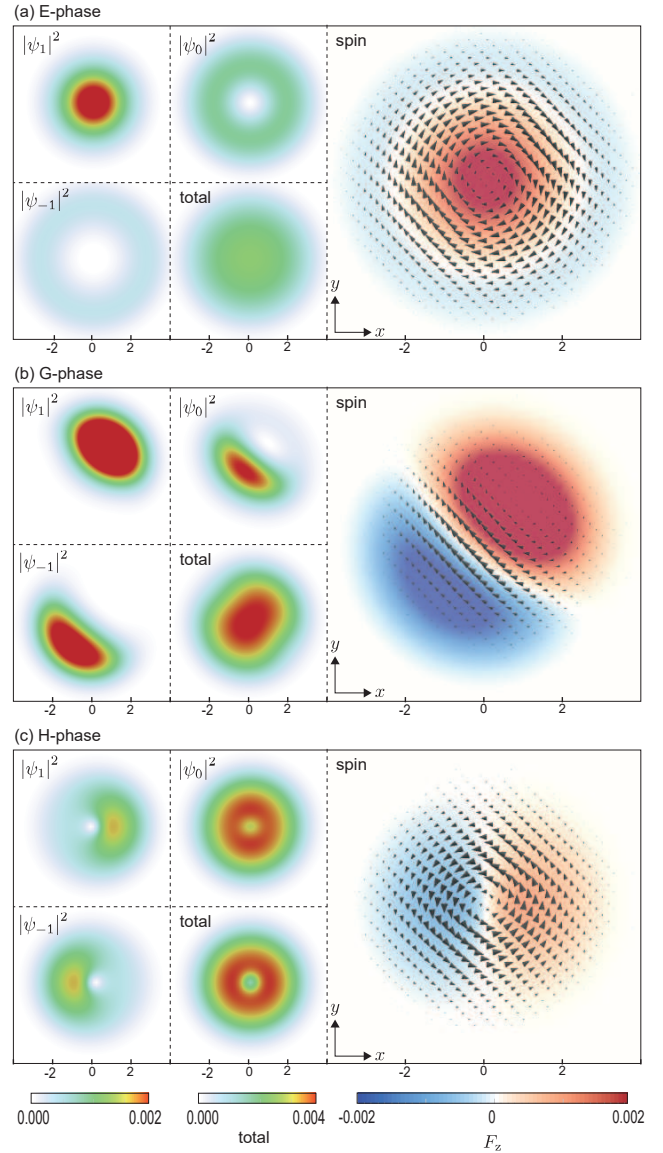}
\caption{(color online) Typical density distribution and spin texture of
the system for
(a) $\textmd{g}_{\rm dd}=300$ and $\textmd{g}_{\rm soc}=0.6$,
(b) $\textmd{g}_{\rm dd}=700$ and $\textmd{g}_{\rm soc}=0.1$,
and (c) $\textmd{g}_{\rm dd}=700$ and $\textmd{g}_{\rm soc}=0$,
corresponding to the states represented in E, G and H phases in
Fig. \ref{f:PD}, respectively.
The arrows in the spin texture represent the transverse spin vector
$(F_{\rm x}, F_{\rm y})$ with background color representing $F_{\rm z}$.
}
\label{f:DDI}
\end{figure}
E-phase is characterized by its axisymmetric density distribution of each
component, where the central region is occupied by component $1$ and outer
regions by components $0$ and $-1$, as shown in Fig.~\ref{f:DDI}(a).
Components $0$ and $-1$ have vorticities $\pm 1$ and $\pm 2$,
respectively.
The spin texture in the $x$-$y$ plane has a single spin vortex at the
center, which is similar to the chiral spin-vortex
state~\cite{Yi,Kawaguchi2}.
This phase exists for strong DDI or weak SOC, and occupies the largest
phase in the ground-state phase diagram in Fig.~\ref{f:PD}.

\begin{figure}[tbp]
\includegraphics[height=5.5cm,width=5.0cm]{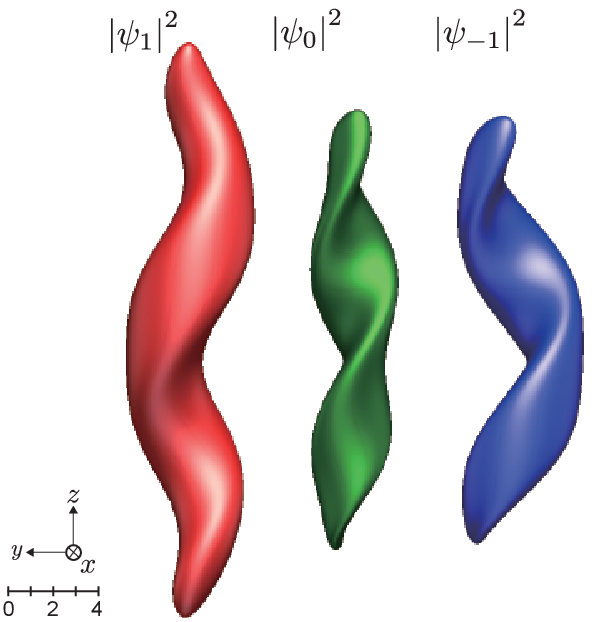}
\caption{(color online) Isodensity surfaces of the three components
$|\psi_1|^2 = 0.001$ (red), $|\psi_0|^2=0.001$ (green), and
$|\psi_{-1}|^2=0.001$ (blue), corresponding to G-phase shown in
Fig.~\ref{f:DDI}.
See the Supplemental Material for a movie showing the 3D
structure~\cite{SM}.
} \label{f:G}
\end{figure}
Finally, we move to another limit of weak SOC and strong DDI.
In this region, there are two phases marked G and H in Fig.~\ref{f:PD}.
The G-phase is shown in Figs.~\ref{f:DDI}(b) and \ref{f:G}.
This phase has a helical structure along the $z$-axis, in which component
$0$ are twined by the other two components, resulting in a double helix of
$F_z > 0$ and $F_z < 0$.
The state shown in Figs.~\ref{f:DDI}(b) and \ref{f:G} has not only the
spin angular momentum but also the orbital angular momentum in the $z$
direction.
For a small $\textmd{g}_{\rm soc}$, $+1$ and $-1$ components are balanced
and the spin and orbital angular momenta disappear.
In the present case of $\textmd{g}_1 = 0$, this state is not the ground
state for $\textmd{g}_{\rm soc} = 0$, while it can be a stationary state.
It is found in Ref.~\cite{Helical} that this state can be the ground
state for $\textmd{g}_1 < 0$ even without SOC.

The density and spin distributions of H-phase are shown in
Fig.~\ref{f:DDI}(c).
The spin texture of this state is similar to that of the polar-core
vortex, i.e., $|\bm{F}| = 0$ on the $z$-axis and the transverse spin
vectors rotate around the core.
However, this state is different from the polar-core vortex, in that the
axisymmetry is broken in $|\psi_{\pm 1}|^2$ and $F_z \neq 0$.

In all the phases demonstrated above, space- and time-reversed states of
the ground states are also the ground states because of the symmetry of
the Hamiltonian, that is, if $\psi_m(\bm{r})$ is a ground state, $(-1)^m
\psi_{-m}^*(-\bm{r})$ is also a ground state.
The rotation about the $z$-axis also does not change the energy.
A-, C-, F-, and H-phases have the space-time reversal symmetry and E-phase
has the rotation symmetry about the $z$-axis.

\begin{figure}[tbp]
\includegraphics[width=6.5cm]{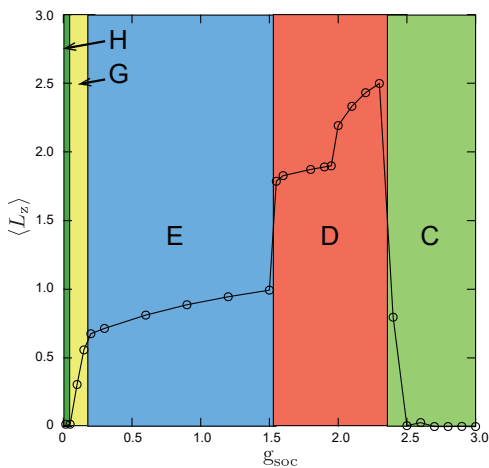}
\caption{(color online) Orbital angular momentum
$\left< L_{\rm z} \right>$ as a function of $\textmd{g}_{\rm soc}$ for
$\textmd{g}_{\rm dd}=700$.
The vertical lines separate the phases and the solid curve is guide to the
eyes.
} \label{f:Lz}
\end{figure}
The $z$-component of the orbital angular momentum,
$\left< L_z \right> = \int \bm r\Psi^\dagger(\bm r)(xp_y - yp_x)\Psi(\bm r)$
is notable in considering our system.
Figure~\ref{f:Lz} shows $\left< L_z \right>$ as a function of
$\textmd{g}_{\rm soc}$ for $\textmd{g}_{\rm dd}=700$ being fixed.
The C- and H-phases scarcely have angular momentum, since there is the
space-time reversal symmetry and $\left< L_z \right>$ is canceled between
$\psi_1$ and $\psi_{-1}$.
The first rapid increase in $\left< L_z \right>$ occurs in the G-phase
($0.06 \lesssim \textmd{g}_{\rm soc} \lesssim 0.17$).
The E-phase also has nonzero $\left< L_z \right>$, since $\psi_0$ and
$\psi_{-1}$ have singly and doubly quantized vortices, respectively.
In the D-phase, $\left< L_z \right>$ changes at $\textmd{g}_{\rm soc}
\simeq 2.0$, since the number of spin vortices changes from two to three.
The maximum of $\left< L_z \right>$ is attained in the D-phase, and
at $\textmd{g}_{\rm soc} \simeq 2.4$, $\left< L_z \right>$ dramatically
decreases and the ground state transforms to the C-phase with an
increase in SOC.

We have also examined the cases of $\textmd{g}_1 \neq 0$, and found that
the B-E phases remained almost unchanged for $|\textmd{g}_1| \sim 0.1
\textmd{g}_0$; the phase boundaries are slightly shifted.
Our main results are thus unchanged for finite $\textmd{g}_1$.

\section{Conclusions}

We have investigated the ground-state structures of a spin-$1$
Bose-Einstein condensate with the 2D Rashba SOC and DDI, confined in a
cigar-shaped trap potential.
Due to the interplay between the 2D-like pattern formation by the Rashba SOC and
the $z$ dependence arising from the long-range DDI, we found a rich
variety of ground-state phases, including the twisted spin vortices.
We systematically explored the parameter space and obtained the
ground-state phase diagram as a function of the strength of the SOC and
DDI, which consists of eight different phases.
For strong SOC and weak DDI, the stripe or plane-wave phase is obtained.
Increasing the DDI, the spin-vortex lattice emerges (B-phase), which form
square pattern due to the long-range nature of the DDI.
In the opposite limit, i.e., for strong DDI and weak SOC, we have two
symmetry broken states (G- and H-phases).
The chiral spin-vortex state (E-phase) is the ground state for a wide
parameter region.
Between B- and E-phases, we found novel spin structures having both SOC
and DDI features, which we call C-and D-phases.
In the C-phase, bunches of spin vortices with opposite directions are
twisted along the $z$-axis, and in the D-phase, a few spin vortices form
helical structures.

Both SOC and DDI couple the internal and external degrees of freedom in a
BEC. Combining such effects, a wide variety of spin textures will be realized.

\begin{acknowledgments}
This work was supported by JSPS KAKENHI Grant Numbers JP16K05505,
JP26400414, and JP25103007, by the key project fund of the CAS for the
``Western Light'' Talent Cultivation Plan under Grant No. 2012ZD02, and by
the Youth Innovation Promotion Association of CAS under Grant
No. 2015334.
\end{acknowledgments}

\end{document}